%% file: main.tex
\documentclass[runningheads]{llncs}
\pdfoutput=1

\usepackage{graphicx}
\usepackage{color}
\usepackage{paralist}
\usepackage{amsmath}
\usepackage{hyperref}
\usepackage{fancyhdr}
\usepackage[multiple]{footmisc}
\usepackage{cite}

\begin{document}
\title{A Transfer Learning  End-to-End  Arabic Text-To-Speech (TTS) Deep Architecture}

\author{Fady K. Fahmy\orcidID{0000-0002-9788-0555} \and
Mahmoud I. Khalil\orcidID{0000-0003-2104-7038} \and
Hazem M. Abbas\orcidID{0000-0001-9128-3111}}

\institute{Ain Shams University, Dept. Computer and Systems Engineering,  Cairo, Egypt
\email{fadykhalaf01@gmail.com, \{mahmoud.khalil,hazem.abbas\}@eng.asu.edu.eg}\
}

\titlerunning{A Transfer Learning  End-to-End  Arabic Text-To-Speech Deep Architecture}

\maketitle              
\vspace{-10pt}
\begin{abstract}
Speech synthesis is the artificial production of human speech. A typical text-to-speech system converts a language text into a waveform. There exist many English TTS systems that produce mature, natural, and human-like speech synthesizers.  In contrast, other languages, including Arabic, have not been considered until recently. Existing Arabic speech synthesis solutions are slow, of low quality, and the naturalness of synthesized speech is inferior to the English synthesizers. They also lack essential speech key factors such as intonation, stress, and rhythm. Different works were proposed to solve those issues, including the use of concatenative methods such as unit selection or parametric methods. However, they required a lot of laborious work and domain expertise. Another reason for such poor performance of Arabic speech synthesizers is the lack of speech corpora, unlike English that has many publicly available corpora\footnote{LjSpeech, \texttt{https://keithito.com/LJ-Speech-Dataset/}}\footnote{Blizzard 2012, \texttt{http://www.cstr.ed.ac.uk/projects/blizzard/2012/phase\_one/}} and audiobooks. This work describes how to generate high quality, natural, and human-like Arabic speech using an end-to-end neural deep network architecture. This work uses just $\langle$ text, audio $\rangle$ pairs with a relatively small amount of recorded audio samples with a total of 2.41 hours. It illustrates how to use English character embedding despite using diacritic Arabic characters as input and how to preprocess these audio samples to achieve the best results. 

\end{abstract}

\input{introduction.tex}

\input{related_work.tex}

\input{architecture.tex}


\input{experiment.tex}

\input{conclusion.tex}

\subsection*{Acknowledgment}
  The authors would like to thank The Bibliotheca Alexandrina for providing the computing resources through their Supercomputing Facility\newline\texttt{(https://hpc.bibalex.org)}.

\bibliographystyle{IEEEbib}
\bibliography{citation}
\end{document}

%% file: introduction.tex
\section{Introduction}
\label{sec:intro}
\vspace{-5pt}
Speech synthesis has been a challenging task for decades. Conventional text-to-speech (TTS) systems are usually made up of several components connected through a pipeline that includes text analysis frontends, acoustic models, and audio synthesis models. Building each component in a conventional TTS system often requires comprehensive domain expertise and a lot of laborious work like feature engineering and annotation. Besides, errors generated by each component propagate to later stages, making it hard to identify the source of the final perceived error.

Researchers have adopted the use of concatenative speech synthesis~\cite{541110, 266409} for years. The idea is based on selecting and concatenating units (phonemes) from a large database to generate intelligible speech. Such units could be any of the following: phones\footnote{distinct speech sound or gesture, regardless of whether the exact sound is critical to the meanings of words}, diphones\footnote{consists of two connected half phones that start in the middle of the first phone and end in the middle of the second phone}, half-phones, syllables, morphemes, words, phrases, or sentences. Generally, the longer the unit, the larger the size of the database that must cover the unit with different prosodies. The drawbacks of concatenative methods for speech synthesis are (a) they need massive databases for large unit size, (b) noise captured while recording units may degrade the quality of synthesized speech since units recorded are represented as it is while synthesizing, and finally (c) the massive amount of labeling and recording.

Statistical parametric speech synthesis based on Hidden Markov Model \newline (HMM) \cite{6495700, 5575397} showed an increase in adoption rate and popularity over time. It solved a lot of problems of concatenative methods such as (a) modeling prosodic variation by modifying HMM parameters, thus solving the problem of large databases, (b) it has proved to have fewer word error rates which lead to better understandably, and (c) it is more robust because the pre-recorded units in unit selection synthesis could be recorded in different environment adhering different noise profiles. The drawbacks of HMM-based synthesis may include (a) requiring a lot of feature engineering and domain expertise, and (b) generated speech sounds more robotic than speech generated by unit selection speech synthesis.

Deep neural network architectures have proved extraordinary efficient  at learning the inherent features of data. WaveNet~\cite{DBLP:journals/corr/OordDZSVGKSK16} is a generative model for generating waveforms based on PixelCNN~\cite{DBLP:journals/corr/OordKK16}. It has outperformed production level parametric methods in terms of naturalness. Still, it has two significant drawbacks: (a) it requires conditioning on linguistic features from an existing TTS system, so it is not a fully end-to-end system, and (b) it synthesizes speech very slowly due to the auto-regressive nature of the architecture.
Deep voice~\cite{DBLP:journals/corr/ArikCCDGKLMRSS17} is another example of deep neural architectures. It has proven high performance, production-level quality, and real-time synthesis. It consists of five stages, namely a segmentation model for locating phoneme boundaries, a grapheme-tophoneme conversion model, a phoneme duration, a fundamental frequency prediction model, and an audio synthesis model. Deep Voice is a step towards a genuinely end-to-end neural network architecture.

With the introduction of end-to-end architectures such as Tacotron~\cite{DBLP:journals/corr/WangSSWWJYXCBLA17}, much laborious work to synthesize speech is alleviated. Such examples for laborious work include feature engineering, and human annotation (although a slight human annotation is needed to prepare the $\langle$ text, audio $\rangle$ pairs for training). Tacotron is a generative text-to-speech model based on a seq-to-seq model with attention mechanism~\cite{DBLP:journals/corr/SutskeverVL14} taking characters as input and producing audio waveforms. Tacotron uses content-based attention~\cite{DBLP:journals/corr/BengioVJS15}, where it concatenates context vector with attention RNN cell output to provide an input to decoder RNNs. Tacotron 2~\cite{DBLP:journals/corr/abs-1712-05884} is a natural evolution of Tacotron. It offers a unified purely neural network approach and eliminates the non-neural network parts used previously by Tacotron, such as the Griffen-Lim reconstruction algorithm to synthesize speech. Tacotron 2 consists of two main components, (a) recurrent seq-to-seq generative model with attention, and (b) a modified Wavenet acting as a vocoder to synthesize speech signal.
Tacotron 2 uses hybrid attention~\cite{NIPS2015_5847} (both location-based and content-based attention).

This paper describes how to use a modified deep architecture from Tacotron 2~\cite{DBLP:journals/corr/abs-1712-05884} to generate mel-spectrograms from Arabic diacritic text as an intermediate feature representation followed by a WaveGlow architecture acting as a vocoder to produce a high-quality Arabic speech. The proposed model is trained using a published pre-trained Tacoron 2 English model using a dataset with a total of 2.41 hours of recorded speech. 

The rest of this paper is organized as follows: Sec. \ref{sec:related} presents a review of related works in the Arabic TTS domain. Sec. \ref{sec:arch} describes the proposed model architecture, including the two main components, feature prediction network and WaveGlow, while Sec. \ref{sec:experiments} introduces the training setup and procedures, issues faced in training, and quantitative and qualitative analysis evaluation of the results. Finally, the paper is concluded in Sec. \ref{sec:conclusion}.

%% file: related_work.tex
\section{Arabic TTS Works}
\label{sec:related}
\vspace{-5pt}
Many works are covering Arabic text-to-speech synthesis to generate a good and human-like speech. In~\cite{45531}, Y. A. El-Imam uses a set of sub-phonetic elements as the synthesis units to allow synthesis of unlimited-vocabulary speech of good quality. The input to the system is an Arabic diacritic spelling or simple numeric expressions.

Abdel-Hamid Ossam et al., in~\cite{Arabic_HMM}, managed to improve the synthesized Arabic speech using an HMM-based approach. They used a statistical model to generate  Arabic speech parameters such as spectrum, fundamental frequency (F0), and phonemes duration. Then, the authors applied a multi-band excitation model and used samples extracted from spectral envelop as spectral parameters.  

Speech synthesis using diacritic text such as~\cite{Rebai2016} has gained a lot of momentum because there is a lack of Arabic diacritic database for speech synthesis. The work discusses two methods to recognize appropriate diacritic marks for Arabic text: a machine learning approach and a dictionary method. This work uses a statistical parameter approach using non-uniform unit size for speech synthesis. It employs variable-sized units, as it has proven to be more effective than using fixed-size units such as phonemes and diphonemes. It partially solves some problems of classical statistical parameter methods. Such issues are speech quality, articulatory effect, and discontinuity effect. This work aimed to build an Arabic TTS system with the integration of diacritization system.

Studying Arabic phonetics~\cite{DBLP:conf/lrec/HalabiW16} for speech synthesis and corpus design is vital to provide a corpus that has excellent coverage of phonetics and phonology.  We have used the corpus generated from~\cite{DBLP:conf/lrec/HalabiW16} in the training phase of the spectrogram prediction network model. We have also used another technique in this work to phonetize diacritic Arabic characters as part of training the spectrogram prediction network.

The work~\cite{zangar:hal-01889917},by Imene Zangar and Zied Mnasri,  uses Deep neural networks (DNN) for duration modeling for Arabic speech synthesis. In this work, the authors compare duration modeling using Hidden Markov Model (HMM) and duration modeling based on deep neural network of different architectures to minimize the root mean square prediction error (RMSE). They concluded that using DNN for modeling duration outperformed HMM-based
modeling from the HTS toolkit and the DNN-based modeling from the MERLIN toolkit.

%% file: architecture.tex
\section{Model Architecture}
\label{sec:arch}
\vspace{-5pt}
Unlike conventional methods for speech synthesis, end-to-end neural network architectures not only alleviate the need for extensive domain expertise and laborious feature engineering, but they also require minimal human annotation. They can be conditioned for any language, gender, or sentiment. Conventional TTS synthesizers consist of many stages, each trained separately. This can give rise to making each component's error cascade to later stages. End-to-end architectures are structured as a single component and thus can become more robust. 

\begin{figure}[t]
\centering
\vspace{-15pt}
\setlength{\abovecaptionskip}{-2pt}
\setlength{\belowcaptionskip}{-10pt}
\includegraphics[width=0.85\textwidth, keepaspectratio]{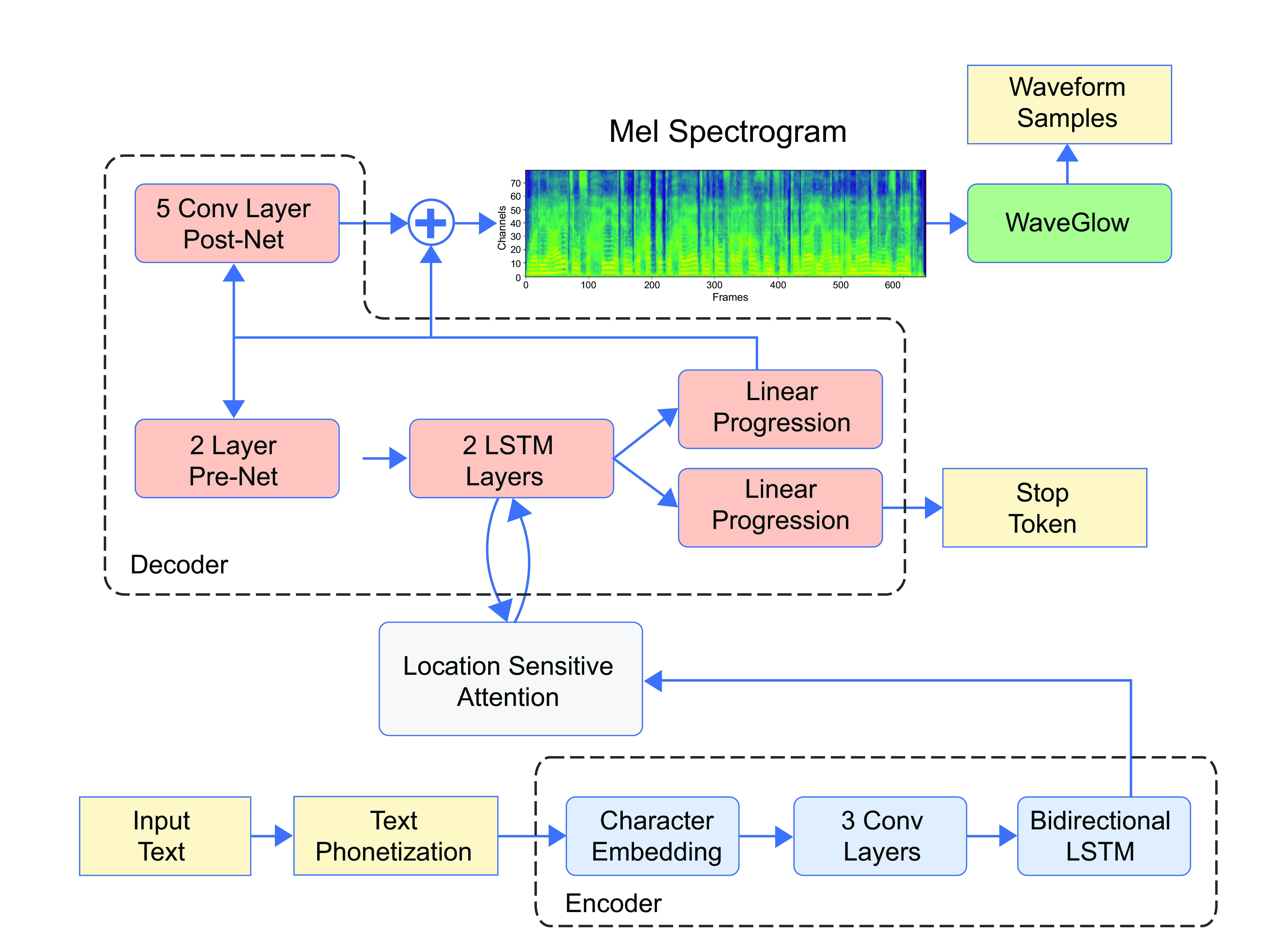}
\caption{Block diagram of the spectrogram prediction network with WaveGlow, it takes diacritic Arabic characters as input and produces audio waveform as output~\cite{DBLP:journals/corr/abs-1712-05884}.}
\label{TTSArchitecture}
\end{figure}

In this work, a slightly modified model that is described in~\cite{DBLP:journals/corr/abs-1712-05884} is adopted where the \emph{Wavenet} part is replaced with a flow-based implementation of \emph{Waveglow}~\cite{DBLP:journals/corr/abs-1811-00002}. Hence, the proposed model shown in Fig. \ref{TTSArchitecture} consists of two components:  
\begin{enumerate}
\item A sequence-to-sequence architecture spectrogram prediction network, with attention which takes a diacritic Arabic text as input and predicts the corresponding mel-spectrogram as output.
\item A flow-based implementation of WaveGlow which takes the mel-spectrograms as input and generates a time-domain waveform of the input text.
\end{enumerate} 
There are many advantages of using mel-frequency spectrograms\footnote{A spectrogram is a visual representation of the spectrum of frequencies of a signal as it varies with time} as an intermediate feature representation between spectrogram prediction network and WaveGlow. They include  
\begin{enumerate}[(a)]
\item mel-spectrograms can be computed easily from time domain waveforms, making it easy to train each of the two components separately.
\item they are easier to train compared to waveforms as they are phase invariant and thus training can be done using simple loss functions such as squared loss.
\item mel-frequency spectrograms are related to linear-frequency spectrograms. One can obtain a mel-frequency spectrogram from a linear-frequency spectrogram by converting the frequency axis to log scale and the "colour" axis, the amplitude, to decibels. 
\item Mel-frequency spectrograms use mel-frequency scale, they can emphasize details on lower frequencies, which is essential for speech naturalness. It also gives less attention to higher frequencies which are not critical for human perception.
\item It is straightforward for WaveGlow to be conditioned on mel-frequency spectrogram to generate a good quality speech.
\end{enumerate}

\subsection{Spectrogram Prediction Network}
As shown in Fig. \ref{TTSArchitecture}, the spectrogram prediction network is a sequence-to-sequence architecture. It consists of an encoder that creates an internal representation of the input signal, which is fed to the decoder to generate the predicted mel-spectrogram. The encoder is made of three parts: character embedding, three convolution layers, and bidirectional LSTM. It takes character sequence as input and produces a hidden feature vector representation. The decoder is made of a two-layer LSTM network, two-layer pre-net, five Conv-layer post-net, and linear progression. It consumes the hidden feature vector representation produced by the encoder and generates the mel-spectrograms of given input characters. Since the diacritic Arabic text is used as an input, a text phonitization block is employed to transform the Arabic characters to another Unicode character set. The following block in the architecture is an embedding layer (512-dimensional vector) which represents each character symbol numerically. The output of the embedding layer is fed to three convolutional layers, each of 512 filters of dimension $5 \times 1$ to span five characters and model long-term contexts (N-gram). Each convolutional layer is followed by batch normalization~\cite{DBLP:journals/corr/IoffeS15} and a ReLU activation~\cite{NIPS2017_6662}.
Tensors produced by the convolutional blocks are fed to bi-directional LSTM of 512 units (256 in each direction). Forward and backward results are concatenated to generate encoded features to be supplied to the decoder.

\begin{figure}[!b]
\vspace{-15pt}
\setlength{\abovecaptionskip}{-2pt}
\setlength{\belowcaptionskip}{-10pt}
\includegraphics[width=0.9\textwidth, keepaspectratio]{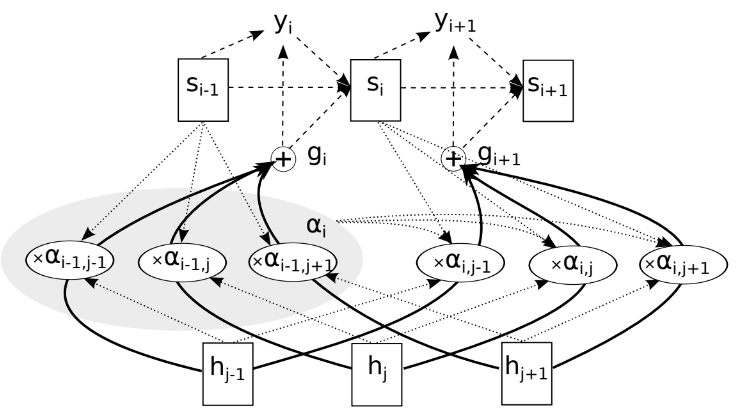}
\centering
\caption{Hybrid attention mechanism used in spectrogram prediction network~\cite{NIPS2015_5847}}
\label{Attention}
\end{figure}

Spectrogram prediction network uses a hybrid attention model described in~\cite{NIPS2015_5847}. The reason why an attention mechanism is necessary for the spectrogram prediction network is solving long sequence problems (long character sequence), as it is hard for encoder-decoder architecture without attention to memorize a very long input sequence. Accordingly, the performance of the architecture without attention mechanism will eventually deteriorate with long sequences. Attention mechanism solves the problem of long sequences by attending on a part of the sequence (using attention weights) just like what human does when trying to figure out a long sequence. As shown in Fig. \ref{Attention}. At each decoder step, to form the context vector and update the attention weights attention uses the following:
\begin{inparaenum}[(a)]
\item the projection of the previous hidden state of decoder RNN's network onto a fully connected layer,
\item the projection of the output of the encoder data on a fully connected layer, and
\item the  additive attention weights.
\end{inparaenum} 
The context vector $C_i$ is computed by multiplying the encoder outputs, $h_j$,  and  the attention weights, $\alpha_{ij}$, as in  Eq. \ref{first} 
\begin{equation}
    c_i = \sum_{j=1}^{T_x}{\alpha_{ij}h_j}
    \label{first}
\end{equation}
\begin{equation}
    \alpha_{ij} = \dfrac{\exp{(e_{ij})}}{\sum_{k=1}^{T_x}\exp{(e_{ik})}}
    \label{second}
\end{equation}
\begin{equation}
    e_{ij} = \mathbf{w}^T \tanh{(W_{s_{i-1}}+V_{h_i}+\mathbf{b})}
    \label{third}
\end{equation}
where $\alpha_{ij}$ is attention weight, 
and $e_{ij}$ is an energy function. 
$W$ and $V$ are matrices, while $\mathbf{w}$ and $\mathbf{b}$ are vectors and they are all trainable parameters.

The output of the decoder layer is then fed to pre-net, which consists of two fully connected layers of 256 hidden ReLU units, then passed through 2 uni-directional LSTM of 1204 units. The concatenation of LSTM output and context vector is projected to a linear transformation to predict mel-spectrogram, which is passed to a five-layer post-net. A scaler (stop token) is calculated in parallel by projecting concatenation of context vector with the decoder LSTM output and passing them through a sigmoid activation to predict when to stop generating speech at inference time. 
Mel-spectrograms are computed using 50ms frame hub, and a "han" window function. 

All convolutional layers are regulated using dropout~\cite{JMLR:v15:srivastava14a}, while LSTM layers are regulated using zoneout~\cite{DBLP:journals/corr/KruegerMKPBKGBL16}.

\subsection{WaveGlow Vocoder}
\vspace{-5pt}
WaveGlow is a flow-based generative network that combines insights from glow ~\cite{kingma2018glow} and Wavenet. According to the authors of ~\cite{DBLP:journals/corr/abs-1811-00002}, it generates speech with quality as good as the best open-source implementations of WaveNet. However, it is much faster as it is not auto-regressive and could fully utilize GPUs. It is trained alongside with the spectrogram prediction network using the original mel-spectrograms as an input and the audio clips as the output. WaveGlow can be easily conditioned on mel-spectrograms to generate high-quality waveforms. 
\begin{figure}[t]
\centering
\vspace{-10pt}
\setlength{\abovecaptionskip}{-2pt}
\setlength{\belowcaptionskip}{-10pt}
\includegraphics[width=0.6\textwidth, keepaspectratio]{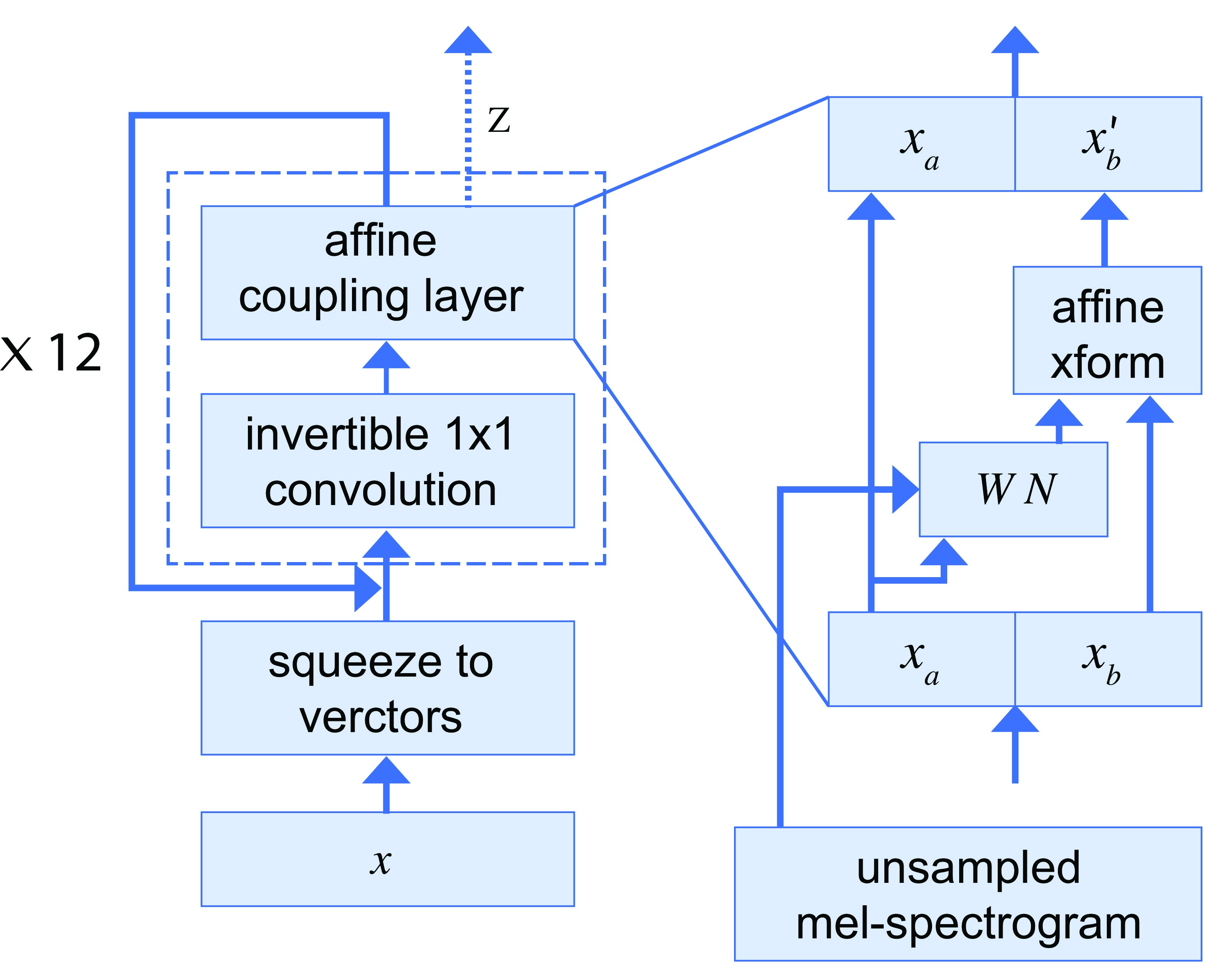}
\caption{Block diagram of WaveGlow Vocoder~\cite{DBLP:journals/corr/abs-1811-00002}, it takes a spectrogram as input and produces an audio waveform.} \label{WaveGlow}
\end{figure}
\label{sec:c2f}

The forward path, as shown in Fig. \ref{WaveGlow}, takes a group of eight samples as a vector as in~\cite{kingma2018glow}, then passes the output into twelve steps of flow, each step consisting of $1 \times 1$ convolution followed by an affine coupling layer. The affine coupling layer acts as an invertible neural network~\cite{DBLP:journals/corr/DinhSB16}. Half of the channels are used as inputs, while the block, WN,  can be any transformation. The coupling layer preserves invertibility for the overall network. Invertible $1 \times 1$ convolution is added before the affine layer to mix information between channels. the weights $W$ of the invertible convolution are initialized to be orthogonal\footnote{orthogonal matrix is a square matrix whose columns and rows are orthogonal unit vectors} and thus they are also invertible.

%% file: experiment.tex
\section{Experimental Results and Analysis}
\label{sec:experiments}
\vspace{-5pt}
\subsection{Training Setup}
\label{sec:Training Setup}
We have trained the Spectrogram prediction network on Nawar Halabi's Arabic Dataset\footnote{\texttt{http://en.arabicspeechcorpus.com/}} \cite{DBLP:conf/lrec/HalabiW16}, which contains about 2.41 hours of Arabic speech, a total of 906 utterances, and 694556 frames. The dataset consists of $\langle$text, audio$\rangle$ pairs. The input text is diacritic Arabic characters, while the output is a 16-bit 48 kHz PCM audio clip is with a bit-rate of 768 kbps. Since the dataset is relatively small, it is split into a 95\% training set and a 5\% validation set.
The training was executed on a supercomputing environment\footnote{\texttt{https://hpc.bibalex.org/}}. 

\begin{figure}[t]
\centering
\vspace{-15pt}
\setlength{\abovecaptionskip}{-2pt}
\setlength{\belowcaptionskip}{-10pt}
\includegraphics[width=0.9\textwidth, height=3cm]{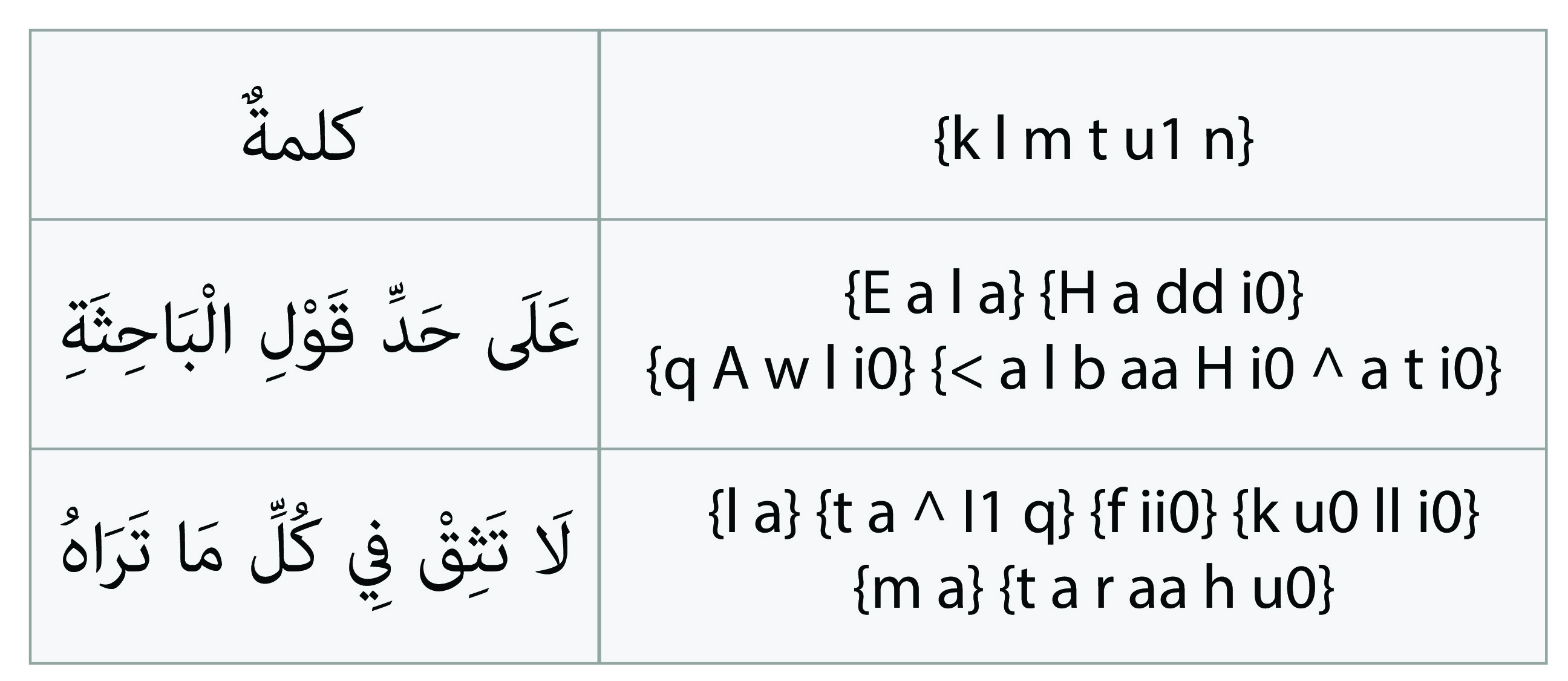}
\caption{Phonitization examples. The left side of the graph represents diacritic Arabic words, while the right side represents the corresponding Unicode character symbols. } \label{fig:phonitization examples}
\end{figure}

\begin{figure}[t]
\centering
\vspace{-8pt}
\setlength{\abovecaptionskip}{-2pt}
\setlength{\belowcaptionskip}{-10pt}
\includegraphics[width=0.7\textwidth, keepaspectratio]{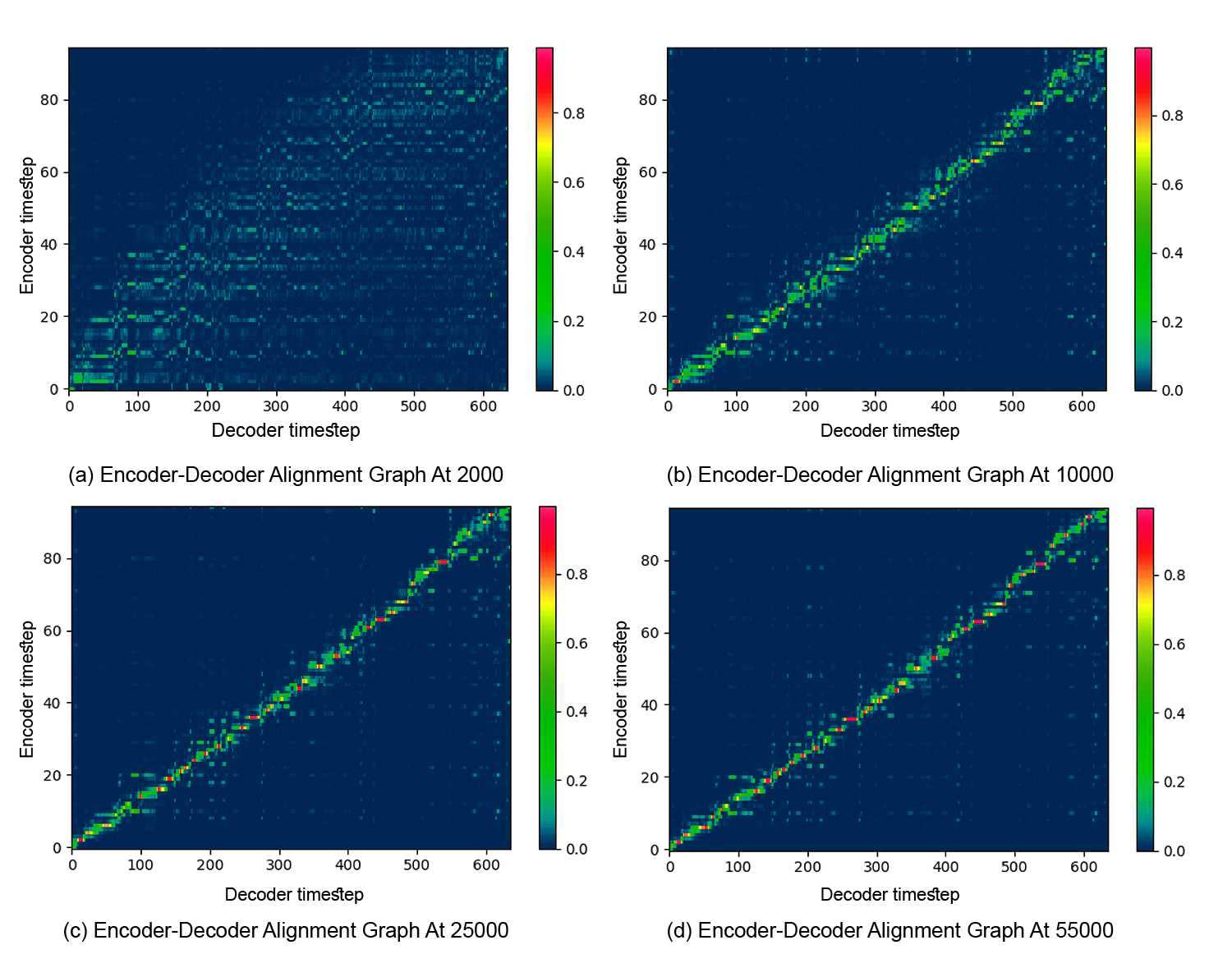}
\caption{Alignment Graphs at different steps of training} \label{Fig:Alignment}
\end{figure}

The spectrogram prediction network was trained separately using diacritic Arabic characters as input and original mel-spectrograms at the decoder side as the target. Because of the small dataset size, we were not able to learn character embedding, nor the attention between encoder and decoder perfectly. Also, the quality of the generated speech was poor. As a result, we utilized transfer learning from English by (a) transforming diacritic Arabic words into English characters using an open-source phonitization algorithm\footnote{\texttt{https://github.com/nawarhalabi/Arabic-Phonetiser}}, refer to Text Phonitization in Fig. \ref{TTSArchitecture}, phonitization examples at Fig. \ref{fig:phonitization examples}, (b) using a pre-trained English model\footnote{\texttt{https://drive.google.com/file/d/1c5ZTuT7J08wLUoVZ2KkUs\_VdZuJ86ZqA/view}} with the learned English character embedding to be able to fully train the attention mechanism. The audio training clips have been down-sampled to 22050 Hz in to employ the same audio parameters as those in the open-source implementation\footnote{\texttt{https://github.com/NVIDIA/tacotron2}} (trained on LJSpeech dataset) such as the hop length and the filter length.  Silence moments (below 60 decibels) of each training sample were removed using a frame size of 1024 and a hop size of 256, which greatly helped to align the attention graph shown in Fig. \ref{Fig:Alignment}.

Other training parameters are: a batch size of 8 on 2 GPUs, Adam optimizer~\cite{DBLP:journals/corr/KingmaB14} with $\beta_1$ = 0.9, $\beta_2$ = 0.999, and $\epsilon = 10 ^ {-6}$,  a constant learning rate of $10 ^ {-3}$, and $L_2$ regularization with weight $10^{-6}$. A training epoch took, on average, about 15 minutes while only about 2 seconds were needed to generate a waveform.

\subsection{Analysis}
\label{sec:Evaluation}
For quantitative analysis, both training and validation losses were assessed as metrics. Simple mean square error loss (MSE) between predicted and target mel-spectrograms was calculated. 

For qualitative analysis, the attention alignment graph was used as a metric. The Attention alignment graph is an indication of how the decoder is attending correctly to encoder input. Encoder reads input step-by-step and produces status vectors. Decoder reads all status vectors and produces audio frames step-by-step. A good alignment simply means: An ``A`` sound generated by the decoder should be the result of focusing on the vector generated by the encoder from reading ``A`` character. The diagonal line is the result when audio frames are generated  by focusing (paying attention) on the correct input characters. Fig. \ref{Fig:Alignment} shows that the spectrogram prediction network was continually improving in learning attention throughout the training process. It helped in eliminating some pronunciation errors as well as removing some pauses in the generated speech. Our model started to pick up alignment after about 40 epochs of training.

Further qualitative analysis was carried out by using human ratings similar to Amazon's Mechanical Turk\footnote{ \texttt{https://www.mturk.com/}}. We used a pre-trained model of WaveGlow\footnote{\texttt{https://drive.google.com/file/d/1rpK8CzAAirq9sWZhe9nlfvxMF1dRgFbF/view}} to infer ten  randomly selected samples of spoken sentences. Each sample is rated by 26 raters on a scale from 1 to 5 with a step of 0.5 to calculate a subjective mean opinion score (MOS) for audio naturalness. Each evaluation is conducted independently from each other. Table \ref{MOS table} compares the proposed architecture with other architectures samples from~\cite{ArabicSpeechComparison} such as concatenative methods with HMMs and Tacotron with the Griffin-Lim algorithm as a synthesizer. Fig. \ref{fig:Result} shows the detailed raters' review for each of the test samples where each entry is the sum of all 26 rates divided by the number of raters (26).

\begin{table}
\setlength{\belowcaptionskip}{0pt}
\centering
\caption{MOS evaluation for different system architectures.}\label{MOS table}
\begin{tabular}{|l|l|}
\hline
System Architecture & MOS\\
\hline
Concatenative methods with HMMs & 3.89\\
\hline
Tacotron 1 with Griffin-Lim algorithm & 4.02\\
\hline
Tacotron 2 with WaveGlow (proposed) & 4.21\\
\hline
\end{tabular}
\end{table}

\begin{figure}[t]
\setlength{\abovecaptionskip}{-2pt}
\setlength{\belowcaptionskip}{-15pt}
\centering
\includegraphics[width=0.6\textwidth,keepaspectratio]{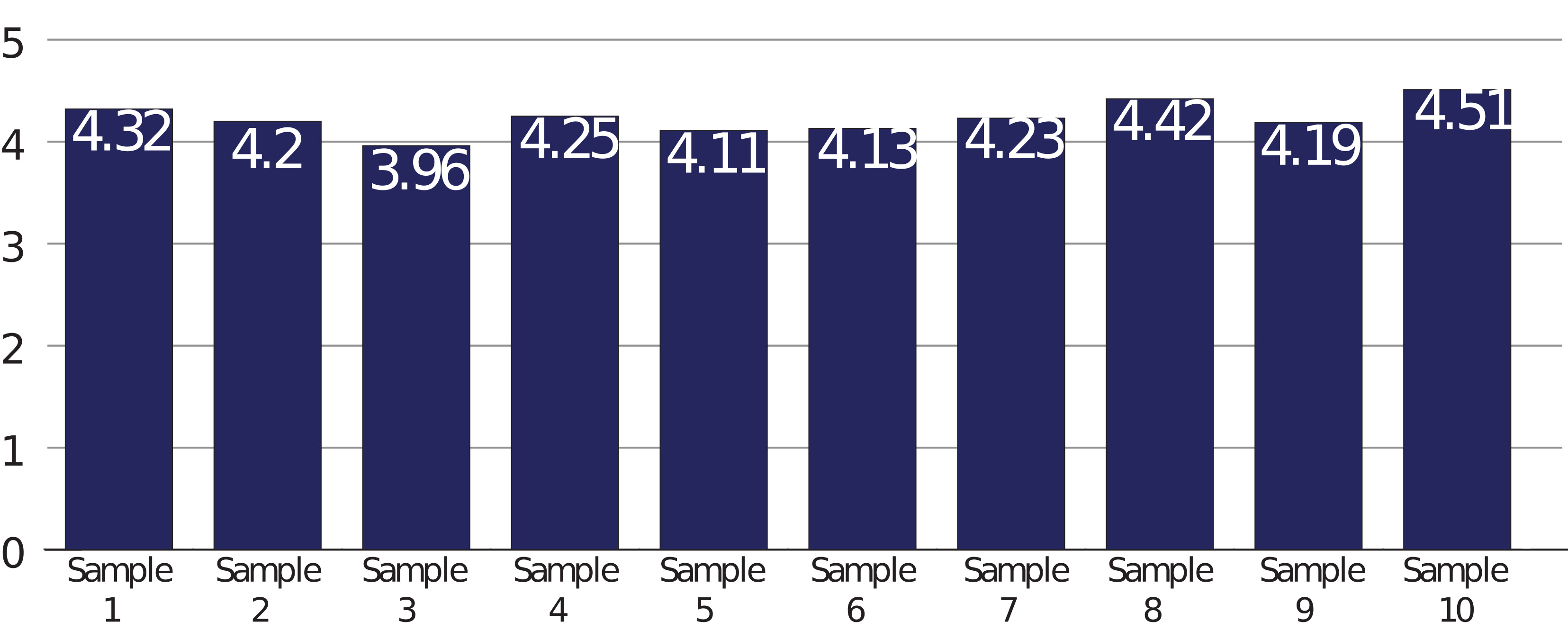}
\caption{Human judgement over ten randomly selected test samples.} \label{fig:Result}
\end{figure}

%% file: conclusion.tex
\section{Conclusions and Future Work}
\label{sec:conclusion}
\vspace{-5pt}
This paper describes how to use the Tacotron 2 architecture to generate intermediate feature representation from Arabic diacritic text using a pre-trained English model and a total of 2.41 hours of recorded speech, followed by WaveGlow as a vocoder to synthesize high-quality Arabic speech.
It also shows the viability of how to apply transfer learning from English text-to-speech to Arabic text-to-speech successfully in spite of the fact that the two languages are quite different in terms of character level embedding and language phonemes. It also describes how to preprocess audio speech training data to gain a plausible generated speech.

There are many possible future enhancements for this work. They may include integrating Arabic diacrtizer, which will reduce the amount of manual work needed to diacrtise a given Arabic text. Another possible enhancement  is to model speech prosody (intonation, stress, and rhythm) for expressive and more human-like speech. Modeling prosody could be done using an architecture similar to Tacotron but with additional neural networks to embed prosody into the  encoded text before encoding the information using the same sequence-to-sequence architecture. And last but not least, using a much larger dataset to train the model which will generally produce  more plausible speech quality.